\documentclass[letter,traditabstract,longauth]{aa} 
\usepackage{graphicx}
\usepackage{natbib}
\bibpunct{(}{)}{;}{a}{}{,}
\def\tCO {\element[][13]{C}O}
\def\CO {\element[][12]{C}O}
\def\CI {[\ion{C}{I}]}
\def\CII {[\ion{C}{II}]}
\def\NII {[\ion{N}{II}]}

\def\GO {$G_0$}
\def\tnx {\footnote{Herschel is an ESA space observatory with science instruments provided by European-led Principal Investigator consortia and with important participation from NASA.}}
\usepackage{txfonts}
\begin{document}
%
\title{Excitation of the molecular gas in the nuclear region of M\,82}

   \author{A.F. Loenen\inst{1}\thanks{\email{loenen@strw.leidenuniv.nl} } \and P.P. van der Werf \inst{1} \and R. G\"usten\inst{2} \and
     R. Meijerink\inst{1} \and F.P Israel\inst{1} \and M.A. Requena-Torres\inst{2} \and\\
     S. Garc\'\i a-Burillo\inst{3} \and 
     A.I. Harris\inst{4} \and 
     T. Klein\inst{2} \and
     C. Kramer\inst{5} \and 
     S. Lord\inst{6} \and 
     J. Mart\'\i n-Pintado\inst{7} \and
     M. R\"ollig\inst{8} \and
     J. Stutzki\inst{8} \and\\
     R. Szczerba\inst{9} \and
     A. Wei\ss\inst{2}\and
     S. Philipp-May\inst{2}\and 
     H. Yorke\inst{10}\and
     E. Caux\inst{11,12}\and
     B. Delforge\inst{13}\and
     F. Helmich\inst{14}\and
     A. Lorenzani\inst{15}\and\\
     P. Morris\inst{16}\and
     T.G. Philips\inst{17}\and
     C. Risacher\inst{14}\and
     A.G.G.M. Tielens\inst{1}
}

   \institute{Leiden Observatory, Leiden University, P.O. Box 9513, NL-2300 RA Leiden, Netherlands \and
     Max-Planck-Institut f\"ur Radioastronomie, Auf dem H\"ugel 69, D-53121 Bonn, Germany  \and 
     Observatorio Astronomico Nacional, Apdo. 1143, E-28800 Alcal\'a de Henares (Madrid), Spain \and 
     Department of Astronomy, University of Maryland, College Park, MD 20742-2421, USA \and 
     Instituto Radioastronom\'{i}a Milim\'{e}trica (IRAM), Av. Divina Pastora 7, 
     Nucleo Central, E-18012 Granada, Spain \and 
     NASA Herschel Science Center, California Institute of Technology, M.S. 100–22, Pasadena, CA 91125 USA  \and 
     LAM, CAB-CSIC/INTA, Ctra de Torrej\'on a Ajalvir km 4, 28850 Torrej\'on de Ardoz, Madrid, Spain \and 
     I.Physikalisches Institut der Universit\"at zu K\"oln, Z\"ulpicher Str. 77, D-50937 K\"oln, Germany\and 
     N. Copernicus Astronomical Center (NCAC), Rabianska 8, P-87100 Torun, Poland \and 
     Jet Propulsion Laboratory, M/S 169-506, 4800 Oak Grove Drive, Pasadena, CA 91109, USA \and
     Institute Centre d'etude Spatiale des Rayonnements, Universite de Toulouse [UPS], 
     31062 Toulouse Cedex 9, France \and
     CNRS/INSU, UMR 5187, 9 avenue du Colonel Roche, 31028 Toulouse Cedex 4, France \and 
     Institute Laboratoire d'Etudes du Rayonnement et de la Matière en Astrophysique, UMR 8112 CNRS/INSU, OP, ENS, 
     UPMC, UCP, Paris, France and LERMA, Observatoire de Paris, 61 avenue de l'Observatoire, 75014 Paris, France \and 
     SRON Netherlands Institute for Space Research, Landleven 12, 9747 AD Groningen \and
     Osservatorio Astrofisico di Arcetri-INAF- Largo E. Fermi 5 I-50100 Florence, Italy\and
     Infrared Processing and Analysis Center, California Institute of Technology, MS 100-22, Pasadena, CA 91125 USA\and
     California Institute of Technology, Cahill Center for Astronomy and Astrophysics 301-17, Pasadena, CA 91125 USA}

   \date{Received June 1, 2010; accepted June 25, 2010} 
   \authorrunning{Loenen et al.}  

   \abstract{ We present high-resolution HIFI spectroscopy of the
     nucleus of the archetypical starburst galaxy M\,82. Six
     \CO\ lines, 2 \tCO\ lines and 4 fine-structure lines have been
     detected.  Besides showing the effects of the overall velocity
     structure of the nuclear region, the line profiles also indicate
     the presence of multiple components with different optical
     depths, temperatures, and densities in the observing beam.  The
     data have been interpreted using a grid of PDR models. It is
     found that the majority of the molecular gas is in low density
     ($n=10^{3.5}$\,cm$^{-3}$) clouds, with column densities of
     $N_{\rm H}=10^{21.5}$\,cm$^{-2}$ and a relatively low UV
     radiation field (\GO = $10^2$). The remaining gas is
     predominantly found in clouds with higher densities
     ($n=10^5$\,cm$^{-3}$) and radiation fields (\GO = $10^{2.75}$),
     but somewhat lower column densities ($N_{\rm H}=10^{21.2}$
     cm$^{-2}$).  The highest $J$ CO lines are dominated by a small
     (1\% relative surface filling) component, with an even higher
     density ($n=10^{6}$\,cm$^{-3}$) and UV field (\GO =
     $10^{3.25}$). These results show the strength of multi-component
     modelling for interpretating the integrated properties of
     galaxies.}
 
   \keywords{Galaxies: individual: M\,82 -- Submillimeter: ISM -- ISM: molecules -- 
     Galaxies: ISM -- Galaxies: starburst}

   \maketitle

\section{Introduction}

M\,82 is one of the best studied starburst galaxies in the local
universe.  Its short distance \citep[3.9\,Mpc,][]{1999ApJ...526..599S}
makes it a superb candidate for detailed studies of the physical
processes related to star formation and their effects on the galaxy.
The emission from the interstellar medium (ISM) is an excellent tool
for diagnosing the physical and chemical properties of the
star-forming environment.  Over the past decades, M\,82 has been
studied in many atomic and molecular species. These observations show
a complex environment where multiple components with different
excitation, temperatures, densities, and filling factors coexists
\citep[e.g.,][]{1992A&A...265..447W, 1996ApJ...465..703L,
  2000A&A...358..433M, 2001A&A...365..571W, 2003ApJ...587..171W,
  2007ApJ...664L..23S, 2008A&A...492..675F}.

In this paper we present observations of the nucleus of M\,82 using
the Heterodyne Instrument for the Far Infrared (HIFI, De Graauw et
al., 2010) on board of the ESA {\it Herschel} Space
Observatory\tnx\ as part of the HEXGAL key programme (PI
R. G\"usten). Due to the large spectral coverage available with {\it
  Herschel}, we can observe a large number of lines, enabling 
comprehensive study of the excitation of the different ISM
components. At the same time, the high spectral resolution provided by
HIFI makes it possible to separate different velocity components.  In
this paper, we combine these observations with detailed modelling to
derive the physical conditions and excitation mechanisms of the
nuclear ISM of M\,82.


\section{Observations, reduction, and results}

\begin{figure*}[t]
  \centering
\includegraphics[width=0.9\textwidth]{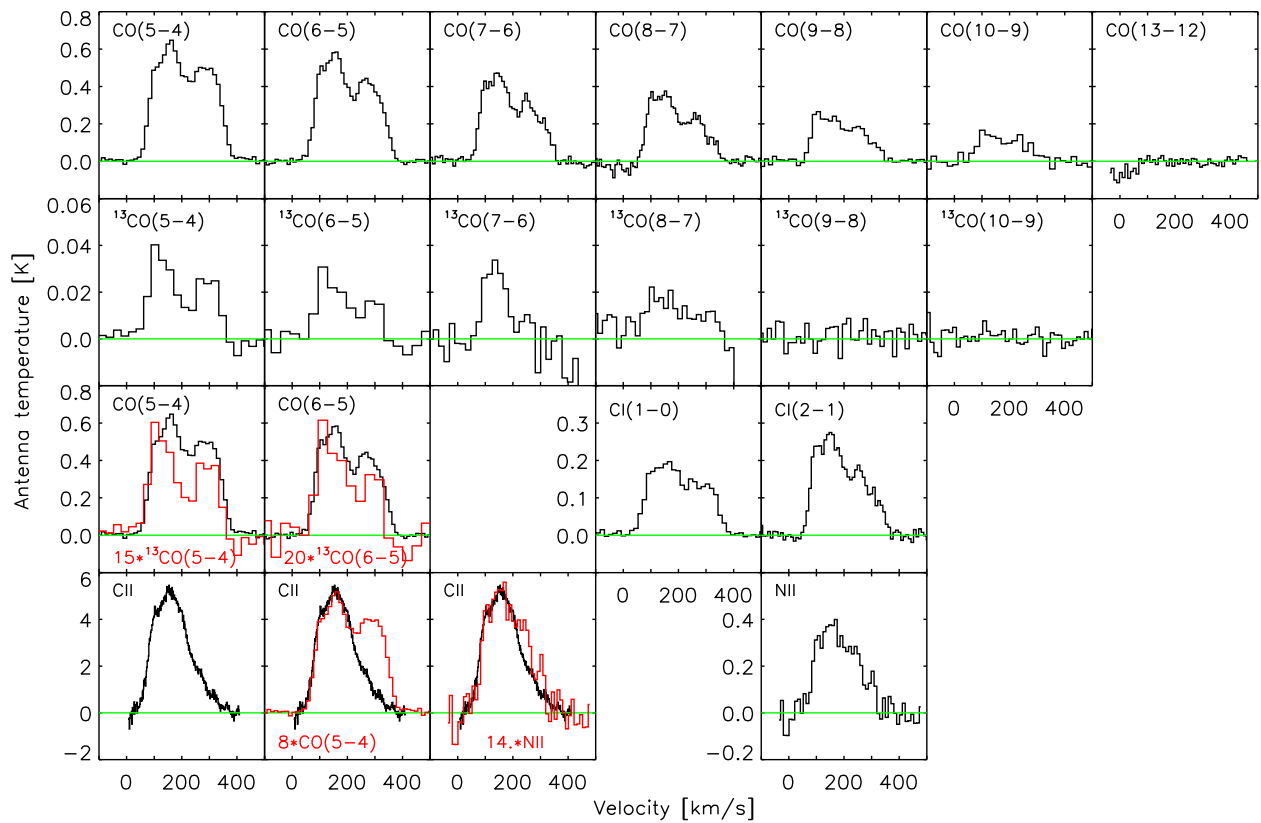}
\caption{Observed spectra. Both polarizations were averaged in these
  spectra. Velocities are heliocentric.}
\label{fig:spectra}
\end{figure*}

The observations were performed in two blocks during April 17-19 and
May 03, 2010. Selected CO transitions and their \tCO\ isotopomers, as
well as fine-structure lines of \CI, \CII, and \NII, were observed
(see Table~\ref{tab:fluxes} and Fig.~\ref{fig:spectra}) in the nuclear
region of M\,82 (RA 09h55m52.22s, Dec 69d40m46.9s J2000).  The
  observations were carried out in dual-beam switch mode with standard
  3\arcmin\ beam-chopping in the ``fast'' mode with chop rates between
  0.2 and 2 Hz to improve stability.  Relevant details of the
  observations are summarized in Table~\ref{tab:fluxes}.

\subsection{Data reduction}
Initial data processing was done using HIPE\footnote{Herschel
  Interactive Processing Environment} version 2.9.  A modified version
of the level 2 pipeline was used that does not time average the data,
in order to inspect the individual subscans in observations that
contain more than one subscan.  After inspection, the subscans were
averaged and the subbands were stitched together.

Further analysis of the data was done using the CLASS
package\footnote{Continuum and Line Analysis Single-dish
  Software:\\http://www.iram.fr/IRAMFR/GILDAS}. The data were binned
until a sufficiently high signal to noise ratio was achieved. Linear
baselines were subtracted, excluding parts of the baseline that
contained lines or seemed affected by instabilities. After this
subtraction, the \tCO\,($5-4$) and \tCO\,($6-5$) spectra showed
residual baseline structure. Since these structures are not seen in
the \CO\,($5-4$) and \CO\,($6-5$) spectra, they were assumed to be
baseline instabilities and  were removed by fitting a 4$^{\rm th}$
order polynomial.

\subsection{Spectra}
Figure~\ref{fig:spectra} shows the resulting spectra, obtained by
averaging both polarizations.  Six CO lines, 2 \tCO\ lines, and all 4
targeted fine-structure lines are detected. Three lines are clear
non-detections: CO\,($13-12$), \tCO\,($9-8$), and \tCO\,($10-9$). The
\tCO\,($7-6$) line can be seen in the spectrum, but the integrated
flux can not be securely determined due to severe baseline
instabilities. The \tCO\,($8-7$) observations also suffer from
baseline problems between $\sim$ 400 and 700\,km\,s$^{-1}$, making it
impossible to determine a good baseline and establish the presence of a
line. Because of the high frequency of the lines, the \CII\ and
\NII\ spectra cover only a narrow range, and therefore a baseline can
only be fit to the outer channels.

\subsection{Line fluxes}
The CO emission in M\,82 has two main velocity components
  \citep[the southwest (SW, v$_{\rm hel}\sim$160\,km\,s$^{-1}$) and
    northeast (NE, v$_{\rm hel}\sim$300\,km\,s$^{-1}$) lobes; see
    e.g., Fig.~\ref{fig:PV} and][]{1992A&A...265..447W}, which can
  also be seen in our data. Therefore, two Gaussian profiles are
  fitted to each spectrum.  The size of the beam varies for the
different frequencies, ranging from 12\arcsec\ for the
\CII\ observations to 44\arcsec\ at the lowest frequency
(\CI\,609\,$\mu$m). To be able to compare the observations to each
other and to the model results, all observations are scaled to the
largest beam. The scaling factor ($\kappa_s$) is calculated by
convolving a 450\,$\mu$m SCUBA map with a resolution of
7\arcsec\ (taken from the SCUBA archive) with Gaussian profiles with
the size of the different beams, to  estimate the flux
contained in those beams. The value of $\kappa_s$ is calculated by
taking the ratio of the integrated 450\,$\mu$m flux for each beam with
the flux contained in the largest beam. This results in values for
$\kappa_s$ between 1 and 0.19 (see Table~\ref{tab:fluxes}). Since the
calibration of HIFI is still preliminary, theoretical predictions
\citep{2006.Kramer.cal} are used to convert the antenna temperatures
into flux densities (see Table~\ref{tab:fluxes}).  After
  applying the beam corrections and the preliminary calibration, the
  final CO fluxes are very similar (within 10\%) to the fluxes
measured with SPIRE, which are corrected for source-beam coupling
using a 250\,$\mu$m SPIRE map \citep{2010arXiv1005.1877P}.  The final
corrected fluxes are listed in Table~\ref{tab:fluxes}, together with
the 1$\sigma$ uncertainties derived from the fits.  Given the
  uncertainties in the pointing and calibration, we apply an
  additional error to the fluxes, when comparing them to the
  models. This error will increase with frequency; however,
  since the exact frequency dependence is unknown, an average value of
  30\% is used.

\section{Analysis and discussion}

\begin{figure}[t]
\centering
\includegraphics[width=\columnwidth]{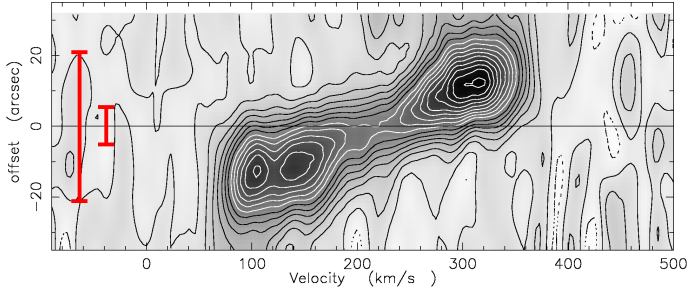}
\caption{Position-velocity diagram of the Nucleus of M\,82, observed
  in \CO($6-5$). Vertical resolution and beam size are
  $\sim$16\arcsec. The two red bars show the largest (44\arcsec) and
  (12\arcsec) smallest beams in our observations.}
\label{fig:PV}
\end{figure}

The spectra in Fig.~\ref{fig:spectra} show well-resolved line
profiles, which provide important boundary conditions for further
a\-na\-ly\-sis. Most lines show the two main velocity components: the
blue-shifted SW and the red-shifted NE emission lobes.  The \NII\ and
\CII\ lines show spectra that are dominated by the blue-shifted
component.  This can occur because the beam at these
frequencies is so small (see beam size bars overlayed on
Fig.~\ref{fig:PV}) that the NE lobe is not covered.  Our
observing position is shown in Fig.~3 of Wei\ss\ et al. (this volume),
where various beam sizes are also indicated, showing the extent to
which the two lobes are covered at various frequencies.

Inspection of the \CO\ line profiles shows that at low $J$ the SW
lobe, peaking at 160\,km\,s$^{-1}$, dominates the blue-shifted
emission component. Going to higher $J$, an additional feature peaking
at 100\,km\,s$^{-1}$ becomes increasingly important and is brighter than
the 160\,km\,s$^{-1}$ peak at $J>8$.  This feature dominates the
\tCO\ lines already at $J=5-4$.  \CO\,($3-2$) and \CO\,($6-5$)
position-velocity diagrams (Fig.~\ref{fig:PV}, data taken from the
JCMT archive) show that this 100\,km\,s$^{-1}$ feature corresponds to
a separate structure in the SW lobe.  The change in contribution of
this feature for different transitions and isotopomers indicates that
it has physical properties different from the neighbouring
160\,km\,s$^{-1}$ feature. This demonstrates that M\,82 contains
star-forming regions with different physical properties, such as optical
depth, temperature, and density, in the observing beam. This
information only becomes available  with the velocity-resolved line
profiles provided by HIFI.

%
\subsection{CO excitation}
The NE and SW lobes, unlike the 100\,km\,s$^{-1}$ spectral feature, do
not show any difference in behaviour. They merely reflect the
kinematics of M\,82 and will both have contributions from different
physical components of the ISM. Because of this and the 
100\,km\,s$^{-1}$ component being too weak in most lines to be fit
separately, we  analyse the CO excitation using the integrated
line strengths, but explicitly allowing for the presence of multiple
excitation components.

\begin{figure}[t]
\centering
\includegraphics[width=\columnwidth]{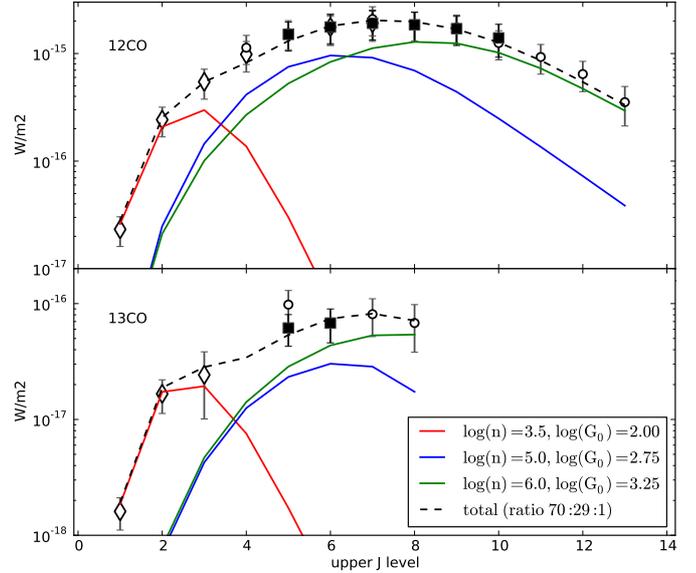}
\caption{Excitation of \CO\ and \tCO. Solid squares represent HIFI
  observations presented in this work, and open diamonds and circles data are
  taken from \cite{2003ApJ...587..171W} and
  \cite{2010arXiv1005.1877P}, respectively. Lines represent the result
  of PDR models. See legend and text for more information on the
  models.}
\label{fig:CO-SLED}
\end{figure}

We  combined our data with ground-based observations of
CO\,($1-0$) to CO\,($4-3$), CO\,($6-5$), CO\,($7-6$), and
\tCO\,($1-0$) to \tCO\,($3-2$) \citep{2003ApJ...587..171W} and the
SPIRE detections of CO\,($4-3$) to CO\,($13-12$), \tCO\,($5-4$),
\tCO\,($7-6$), and \tCO\,($8-7$) \citep{2010arXiv1005.1877P}.
\cite{2003ApJ...587..171W} present separate observations of the two
lobes. Since both fit within our largest beam, we added the fluxes.
The ground-based data have been scaled to match our observations,
using the CO\,($6-5$) and CO\,($7-6$) lines. Both the
\cite{2003ApJ...587..171W} and \cite{2010arXiv1005.1877P} data have
already been corrected for source-beam coupling factors. This resulted
in the two CO excitation diagrams shown in Fig.~\ref{fig:CO-SLED}.

The CO and the fine-structure line ratios are compared to the results
of a large grid of photon-dominated region (PDR) models
\citep{2005A&A...436..397M, 2007A&A...461..793M}. The
\CO/\tCO\ abundance ratio is fixed at a value of 40 \citep[similar to
  the value found by][]{2003ApJ...587..171W}.  The comparison with the
CO lines is shown in Fig.~\ref{fig:CO-SLED}.  The observations can be
best reproduced with a combination of one low and two high-density
components. The low-density component, which dominates in the lowest
CO lines ($J\le3$), is found to have a density of
$n=10^{3.5}$\,cm$^{-3}$ and a UV flux of \GO = $10^2$. This component
represents the extended molecular ISM.  The mid-$J$ CO lines ($4<J<7$)
are mostly produced by clouds that have a density of
$n=10^5$\,cm$^{-3}$ and a radiation field of \GO = $10^{2.75}$. At
transitions higher than 7 (and \tCO\ $J>5$), the line emission comes
from clouds with even higher densities ($n=10^{6}$\,cm$^{-3}$) and
radiation fields (\GO = $10^{3.25}$).  These two models represent the
dense, star-forming molecular clouds. The densest component has
parameters similar to hot cores in the Milky Way. This component has
excitation properties similar to the $100\,$km\,s$^{-1}$ feature
detected in our spectra as can be seen by comparing the \CO\ and
\tCO\,($6-5$) lines in the models.

Since we have \tCO\ lines for both the low- ($J\le3$) and high-density
($J\ge4$) components, it is possible to determine the column
density for both. The low-density component is found to
have a slightly higher column density ($N_{\rm H}=10^{21.5}$\,cm$^{-2}$) than
the high-density component ($N_{\rm H}=10^{21.2}$ cm$^{-2}$). This is
reflected in the \CO/\tCO\ line ratios, which increase from 14.5 for
the $J=1-0$ to 27.2 for the $J=8-7$ transition.  

The relative scaling of the three models provides an estimate of the
relative beam filling of the components and was found to be
70\%:29\%:1\% for the low density, high density low \GO, and high
density high \GO\ components, respectively.

\begin{table*}[t]
\caption{Observation parameters and observed line fluxes, where fluxes include beam correction and calibration.}
\label{tab:fluxes}	   
\centering		       
\begin{tabular}{@{\extracolsep{-1mm}}l r c c c c c c c r r r@{\extracolsep{-1mm}}}	    
\hline\hline		    
\noalign{\smallskip}
\multicolumn{1}{c}{Line}& \multicolumn{1}{c}{Freq}    & $\lambda$ & \multicolumn{1}{c}{Band}    & 
 \multicolumn{1}{c}{t$_{\rm int}$}   &  \multicolumn{1}{c}{Beam}    & \multicolumn{1}{c}{$\kappa_s$} 
 & \multicolumn{1}{c}{$\eta_{MB}$} & \multicolumn{1}{c}{T$_A$$\rightarrow$Jy}& 
\multicolumn{1}{c}{Blue} & \multicolumn{1}{c}{Red} & \multicolumn{1}{c}{Total}	   \\
 & \multicolumn{1}{c}{[GHz]} &[ $\mu$m] &  & \multicolumn{1}{c}{[sec]} &\multicolumn{1}{c}{[\arcsec]} & & & 
\multicolumn{1}{c}{[Jy K$^{-1}$]} & \multicolumn{1}{c}{[$10^3$ Jy km s$^{-1}$]} & 
\multicolumn{1}{c}{[$10^3$ Jy km s$^{-1}$]} & \multicolumn{1}{c}{[$10^3$ Jy km s$^{-1}$]} \\
\hline
\noalign{\smallskip}
CO $J=5-4$	           &  576 & 520 & 1b & 250 & 38 & 0.87 & 0.69 & 474 &   44.2$\pm$0.4  &   34.3$\pm$0.4	 &   78.5$\pm$0.6  \\ 
CO $J=6-5$		   &  691 & 434 & 2a & 248 & 33 & 0.77 & 0.68 & 481 &	44.2$\pm$0.8  &   32.5$\pm$0.8	 &   76.6$\pm$1.2  \\
CO $J=7-6$		   &  807 & 372 & 3a & 248 & 27 & 0.61 & 0.68 & 481 &	40.6$\pm$1.6  &   30.5$\pm$1.7	 &   71.2$\pm$2.3  \\
CO $J=8-7$		   &  922 & 325 & 3b & 250 & 25 & 0.55 & 0.68 & 481 &	33.8$\pm$2.6  &   26.4$\pm$2.7	 &   60.2$\pm$3.8  \\
CO $J=9-8$		   & 1037 & 289 & 4a & 306 & 23 & 0.48 & 0.67 & 488 &	31.4$\pm$1.4  &   18.0$\pm$1.4	 &   49.3$\pm$2.0  \\
CO $J=10-9$		   & 1152 & 260 & 5a & 334 & 20 & 0.40 & 0.67 & 488 &	16.9$\pm$3.7  &   19.5$\pm$4.3	 &   36.4$\pm$5.7  \\
\tCO $J=5-4$		   &  551 & 544 & 1a & 314 & 44 & 1.00 & 0.69 & 474 &	 2.0$\pm$0.1  &    1.3$\pm$0.1	 &    3.4$\pm$0.2  \\	 
\tCO $J=6-5$		   &  661 & 453 & 2a & 306 & 33 & 0.77 & 0.68 & 481 &	 2.3$\pm$0.3  &    0.8$\pm$0.2	 &    3.1$\pm$0.4  \\	 
\CI\ $^3P_1-^3P_0$	   &  492 & 609 & 1a & 248 & 44 & 1.00 & 0.69 & 474 &   13.3$\pm$0.6  &    8.2$\pm$0.6	 &    21.5$\pm$0.8  \\ 
\CI\ $^3P_2-^3P_1$	   &  809 & 369 & 3a & 306 & 27 & 0.61 & 0.68 & 481 &   22.2$\pm$1.9  &   20.3$\pm$2.1	 &    42.5$\pm$2.8  \\ 
\CII\ $^2P_{3/2}-^2P_{1/2}$ & 1901 & 158 & 7b & 306 & 12 & 0.19 & 0.63 & 519 & 1051.4$\pm$57.6 & 1262.7$\pm$57.0 &  2314.1$\pm$81.1 \\ 
\NII\ $^3P_1-^3P_0$	   & 1461 & 205 & 6a & 507 & 15 & 0.27 & 0.65 & 503 &   58.0$\pm$39.1 &   66.7$\pm$39.1 &   124.7$\pm$55.3 \\ 
\hline
\end{tabular}
\end{table*}

\subsection{Fine-structure lines}

To further constrain the modelling results, we also considered
fine-structure line ratios. The ratio between the two \CI\ lines is
sensitive to the impinging UV flux
\citep[see][]{2007A&A...461..793M}. Line ratios with \CII\ and
\NII\ may provide additional insight into the ionization balance.
Because of the different shapes of the red component of the \CII\ and
\NII\ lines, we do not use the total integrated flux of those
lines. Instead, the blue component is scaled using the average
total/blue ratio.

The model prediction for the \CI\,609\,$\mu$m/\CI\,369\,$\mu$m line
ratio (0.28) is close to the observed value (0.31). This is surprising
since \CI\,609\,$\mu$m is usually observed to be stronger than
predicted by models or observed in the Milky Way
\cite[e.g.,][]{1994A&A...284L..23W, 2002A&A...383...82I}. That our
prediction is close to the observed value stems from our
multi-component modelling. The low-density component has a ratio that
is higher than observed, and the high-density components have lower
ratios, resulting in an average that is close to the observed value.

The models under-predict the \CII/\CI\ ratios by about a factor of
three higher.  This excess in \CII\ emission likely arises from the
diffuse, ionized ISM, which is confirmed by the similarity of the
\CII\ and \NII\ line shapes of the blue lobe.

To address the uniqueness of our model, a model with only one
high-density component was also attempted. The best fit is found using
a lower density ($n=10^{4.25}$\,cm$^{-3}$) and a high radiation field
(\GO = $10^{3.5}$). Also a much higher surface ratio is needed (ratio
of low- to high-density components 20\%:80\%).  Although this model is
able to fit the \CO\ excitation within the error bars, it
underpredicts the strength of the \tCO\ lines by factors of a
few. Also the predicted \CII/\CI\ ratios are an order of magnitude too
high and, therefore, a two component interpretation, such as done by
\cite{2010arXiv1005.1877P}, is insufficient for modelling the CO
excitation in M82. This conclusion is reinforced by earlier analyses
by e.g., based on large velocity gradient analysis of multiple CO
lines \citep[e.g.,][]{1992A&A...265..447W, 1993ApJ...402..537G,
  2001A&A...365..571W, 2003ApJ...587..171W, 2005A&A...438..533W}

\section{Summary and outlook}

The CO and fine-structure lines provide an excellent tool for determining
the physical parameters of the nuclear ISM of M\,82. The line shapes
reveal distinct velocity components at 100, 160, and 300\,km\,s$^{-1}$.
While the 160 and 300\,km\,s$^{-1}$ features reflect the large-scale
kinematics of M\,82, the 100\,km\,s$^{-1}$ feature shows an excitation
different from the other components, which is only revealed by the
velocity-resolved HIFI data.  The presence of multiple ISM components
with different physical conditions is also found in modelling of the
  CO lines, which shows that three density components are needed to
  explain the observed line fluxes.

The majority of the gas is in low density ($n=10^{3.5}$\,cm$^{-3}$)
clouds, with column densities of $N_{\rm H}=10^{21.5}$\,cm$^{-2}$ and
a relatively low UV radiation field (\GO = $10^2$). The remaining gas
is predominantly found in clouds with higher densities
($n=10^5$\,cm$^{-3}$) and radiation fields (\GO = $10^{2.75}$), but
lower column densities ($N_{\rm H}=10^{21.2}$ cm$^{-2}$).  The highest
$J$ CO lines are dominated by a small (1\% relative surface filling)
component, with an even higher density ($n=10^{6}$\,cm$^{-3}$) and UV
field (\GO = $10^{3.25}$).

Using the unique spectral coverage and resolution offered by HIFI, we
are one step closer to correctly dissecting the intricate and interacting
density components and excitation regimes present in the core of
M\,82. Additional observations (part of the HEXGAL programme) of the
lobes of M\,82 and of other starburst galaxies will  increase
our understanding of such systems even more.

\begin{acknowledgements}

RSz acknowledges support from grant N 203 393334 from Polish MNiSW.
Support for this work was provided by NASA through an award issued by
JPL/Caltech. Data presented in this paper were analysed using "HIPE",
a joint development by the Herschel Science Ground Segment Consortium,
consisting of ESA, the NASA Herschel Science Center, and the HIFI,
PACS, and SPIRE consortia (See
http://herschel.esac.esa.int/DpHipeContributors.shtml).  HIFI has been
designed and built by a consortium of institutes and university
departments from across Europe, Canada, and the United States under the
leadership of SRON Netherlands Institute for Space Research,
Groningen, The Netherlands, and with major contributions from Germany,
France, and the US.  Consortium members are: Canada: CSA, U.Waterloo;
France: CESR, LAB, LERMA, IRAM; Germany: KOSMA, MPIfR, MPS; Ireland,
NUI Maynooth; Italy: ASI, IFSI-INAF, Osservatorio Astrofisico di
Arcetri- INAF; Netherlands: SRON, TUD; Poland: CAMK, CBK; Spain:
Observatorio Astronómico Nacional (IGN), Centro de Astrobiología
(CSIC-INTA). Sweden: Chalmers University of Technology - MC2, RSS \&
GARD; Onsala Space Observatory; Swedish National Space Board,
Stockholm University - Stockholm Observatory; Switzerland: ETH Zurich,
FHNW; USA: Caltech, JPL, NHSC.

\end{acknowledgements}

\bibliographystyle{aa}
\bibliography{15114}

\end{document}